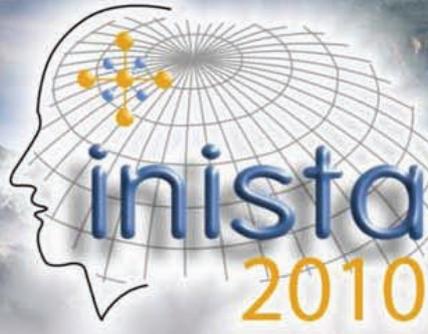

# International Symposium on Innovations in Intelligent SysTems and Applications

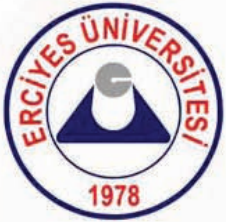 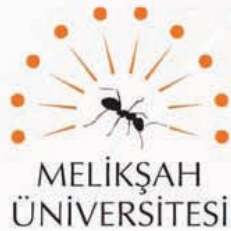 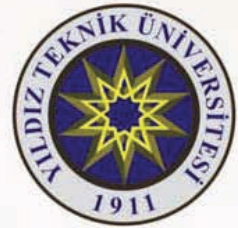

**21 - 24 June 2010**
**Kayseri, TURKEY**

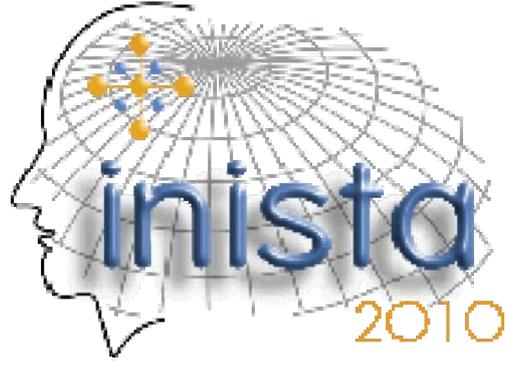

# INISTA 2010

## International Symposium on INnovations in Intelligent SysTems and Applications

**Organized by**

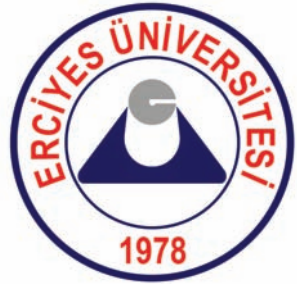

**in cooperation with**

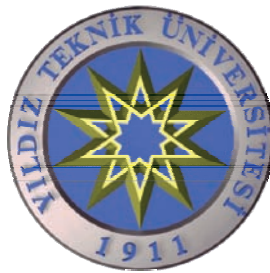 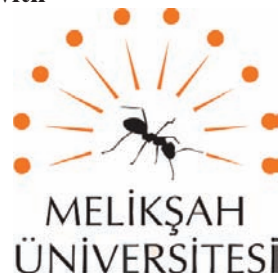

**Technical Co-Sponsors**　　　　　　　　　　**Sponsors**

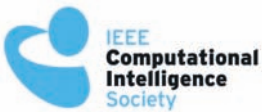 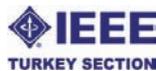 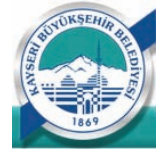 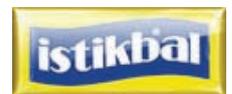

**21-24 June 2010, Kayseri & Cappadocia, TURKEY**

# INISTA 2010

# International Symposium on INnovations in Intelligent SysTems and Applications

**Proceeding Book Name**
International Symposium on INnovations in Intelligent SysTems and Applications
INISTA 2010

**Editors**
Tülay YILDIRIM
Mehmet Emin YÜKSEL
Alper BAŞTÜRK
Enis GÜNAY

**Cover Design**
Nurettin ÜSTKOYUNCU
Rifat KURBAN
Fehim KÖYLÜ
Mehmet YILDIRIM

21-24 June 2010, Kayseri & Cappadocia, TURKEY

# Procedural and Non-Procedural Implementation of Search Strategies in Control Network Programming

Kostadin Kratchanov[1], Emilia Golemanova[2], Tzanko Golemanov[2],
Tuncay Ercan[1] and Burak Ekici[1]

[1]Department of Computer Engineering, Yasar University, Universite Cad. Yagacli Yol No.35-37, Bornova/Izmir, 35100 Turkey
{kostadin.kratchanov, tuncay.ercan}@yasar.edu.tr

[2]Department of Computing, Rousse University, 8 Studentska Street, Rousse, Bulgaria
{EGolemanova, TGolemanov}@ecs.ru.acad.bg

## Abstract

*This report presents the general picture of how Control Network Programming can be effectively used for implementing various search strategies, both blind and informed. An interesting possibility is non-procedural solutions that can be developed for most local search algorithms. A generic solution is described for procedural implementations.*

## 1. Introduction

**Control Network Programming** (**CNP**) evolved as a programming paradigm by combining features from imperative programming, declarative programming, and problem solving with rule-based systems, at the same time substantially extending their potential [1].

### 1.1. Control Network Programming

The program in CNP can be visualized as a finite set of graphs referred to as a **control network** (**CN**). The graphs comprising the CN are called subnets; one of them is identified as the main subnet. Each subnet consists of nodes (states) and arrows. A chain of "primitives" is assigned to each arrow. The primitives may be thought of as elementary actions and are technically user-defined procedures. A subnet may call other subnets or itself. Both subnets and primitives can have parameters and variables. The complete program consists of two main components - the CN and the definitions of the primitives.

The program is a possibly nondeterministic description of a problem. "Executing" the CN means traversing the CN starting from the unique initial node of the main subnet and executing the primitives along the way. This process will successfully finish when the interpreter arrives at a system node *FINISH*. The computation/search strategy is an extended version of backtracking; one of the major enhancements is the possibility to backtrack through a subnet.

CNP allows a convenient combination of procedural and non-procedural features – while the structure of the CN is generally nonprocedural (i.e., can be a simple representation of the problem without specifying any algorithm for solving this problem) and nondeterministic, the primitives used in it are procedural. However, procedural solutions can also be easily programmed.

The syntax and semantics of CNP, and most specifically of the CNP programming language *Spider*, have been described in [1-3]. Representative examples of using CNP for solving various types of problems have been considered in [4]. Programming environments for developing and running CNP applications are available for free download at [5]; the code of all examples from the mentioned publications has also been posted there.

CNP supports powerful means (system options and control states) that give the programmer extensive control over the computation (inference). Using these means the programmer can improve the efficiency of the computation and easily implement various types of heuristic algorithms. These control features are their usage are studied in [6-8].

### 1.2 Purpose of this report

Search algorithms play a fundamental role in Artificial Intelligence. They also attract considerable interest in Algorithm Design, Computability Theory, Operations Research, some areas of Mathematics and Engineering, Robotics, Bioinformatics, and other fields.

Numerous examples of using CNP for implementing search strategies are discussed in [8]. The focus is on employing the built-in tools for dynamic control of the computation process for automatic, non-procedural modeling of certain search strategies.

This report studies the possibilities for programming search algorithms in CNP in a more general framework. Two major implementation techniques are outlined. In the first approach, the classical search algorithms are essentially simulated in CNP – we refer to such implementations as **procedural implementations**. We describe a generic search CNP solution which can be used for modeling many fundamental strategies for uninformed or for heuristic search. Then, for a group of search strategies that we call local search strategies, we show a very different approach where the CN simply describes the problem but includes no explicit procedure for implementing the search process at all; instead, dynamic control system options and control states are used in order to enforce the interpreter to "automatically" perform the desired search strategy. We refer to this type of implementations as **non-procedural implementations**. Finally, certain search strategies require an approach that combines features of both procedural and non-procedural implementations.

The *Spider* code of all examples described here is available at [5]. Typically, each strategy is applied for two particular problems – the road map problem [3,7-10] and the 8-puzzle problem (e.g., [9-11]). The specific map used in the examples is the one shown in Fig.7 of [3] and Fig.3 of [7].

## 2. Procedural Implementation of Search Strategies

This approach can be used for implementing many of the fundamental search strategies, both blind and informed; the strategies we cover are breadth-first, depth-first (leap frogging), uniform-cost, best-first, A*. The strategies use data structures usually called CLOSED and OPEN containing, respectively, the nodes already explored and the nodes at the fringe of the search. Each node may be accompanied by some additional information such as the cost from the initial node to this node, a heuristic evaluation of the cost from this node to a final node, a parent of the node. While CLOSED is theoretically a set, container OPEN is ordered (the order depending on the strategy being implemented).

### 2.1 The generic search algorithm

In order to come up with a CNP solution, we develop first the pseudo-code of a generic search algorithm that generalizes all mentioned strategies. Then we create the corresponding UML activity diagram and (in an almost trivial manner) convert this diagram into a CN. A similar conversion was used in [4] for solving the Selection Sort example.

The generic search algorithm is unique and works for all strategies; the strategy is being chosen through a dialogue with the user. The generic algorithm is based on the observation that all other search strategies under consideration are actually special cases of the A* algorithm. A similar but narrower universality observation can be found in [10,11].

```
procedure Generic_Search;
begin
  OPEN ← [makeInitEntry(START)];
  CLOSED ← [];
  while OPEN <> [] do
  begin
    popOpen(S);
    if final(S) then return(solution)
    else begin
           pushClosed(S);
           findChildren(S);
           pushOpen;
           sortOpen;
         end; {else}
  end; {while}
  return(failure);
end; {Generic_Search}
```

**Figure 1.** The generic search algorithm

```
procedure pushOpen;
begin
for each T=<child,g,h,parent> in the set of the children do
  begin
  case
    CLOSED contains an element <child,g',?,?>:
      begin
        if g < g' then
          begin
            delete <child,g',?,?> in CLOSED;
            add <child,g,h,parent> to OPEN;
          end
      end
    OPEN contains an element <child,g',?,?>:
      if g < g' then
        replace <child,g',?,?> by <child,g,h,parent> in OPEN;
    else:
      add <child,g,h,parent> to OPEN;
  end; {case}
  end; {for each}
end; {pushOpen}
```

**Figure 2.** Procedure pushOpen

The pseudo-code of the generic algorithm is shown in Fig. 1. It uses function *pushOpen* specified in Fig. 2.

We distinguish between a state and a **complete state**. The latter is an ordered quadruple <*state,g,h,parent*> where *g* and *h* are values called **cost** and **heuristic evaluation**, respectively. The last component, *parent* is used for restoring the solution path after finding a solution. OPEN is a set of complete states. At each step of the algorithm it is re-ordered with respect to a strategy-dependent function, *f* referred to as the **total function**.

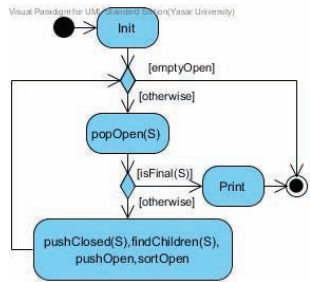

**Figure 3** The generic search algorithm - activity diagram

Function *makeInitEntry(state)* converts *state* into a corresponding complete state *<state,0,?,nil>*. CLOSED is a set of complete states. Procedure *popOpen(S)* assigns its output parameter *S* of type complete state the value of the first element of OPEN, and removes this element from OPEN. The *solution* is typically a path of states. It can be restored from the final state using the *parent* components in CLOSED. It might also contain additional information such as the cost of the solution. Procedure *findChildren(S)* generates the set of all successors of the complete state *S* – the current state. This procedure is problem-dependent. If *S=<state,g,h,parent>* then, for each successor, *newState* of *state*, the cost of *newState* will be calculated as follows: for A* and unified-cost search *newState.g:=state.g+ arc_cost (state,newState)* where *arc_cost(state,newState)* is the cost of the arrow between the two states; for best-first, breadth-first and depth-first search strategies *newState.g:= state.g+1* (i.e., depth). The heuristic evaluation of the child is calculated by a problem-dependent function, *h*. The value *newState.parent* is set to *state*. Procedure *pushOpen* is strategy-independent. Some of the children of *S* are killed; the surviving children are pushed into OPEN. The algorithm of this procedure becomes clear from Fig. 2. Set OPEN is sorted in ascending order according to the values of the following algorithm-dependent total function: for a given complete state *T=<state,g,h,parent>* the value *f(T)* is equal to *g+h* if the strategy is A*, equal to *h* for best-first search, to *g* for the uniform-cost and breadth-first strategies, and to *–g* for depth-first search. We assume here that optimal for both costs and heuristic evaluations means the smallest.

The activity diagram of the generic search algorithm is shown in Fig. 3. The CN of the CNP implementation is illustrated in Fig. 4.

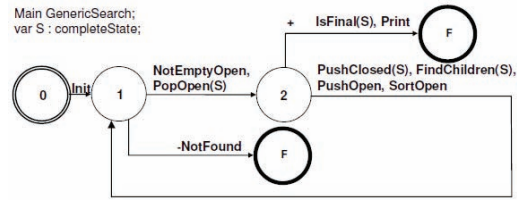

**Figure 4.** CN for the generic search algorithm

Note that the CNP implementation of the generic search algorithm discussed above is universal and covers all the search strategies under consideration. The actual strategy is chosen by the user during an initial dialogue (in which the initial and the final states are also specified). In our implementation, the code of all primitives is generic and does not depend on the strategy. The choice of the strategy made by the user is used in two simple supporting functions – one employed in primitive *findChildren* to calculate *g*, and one in primitive *sortOpen* for calculating *f* (see the explanations above).

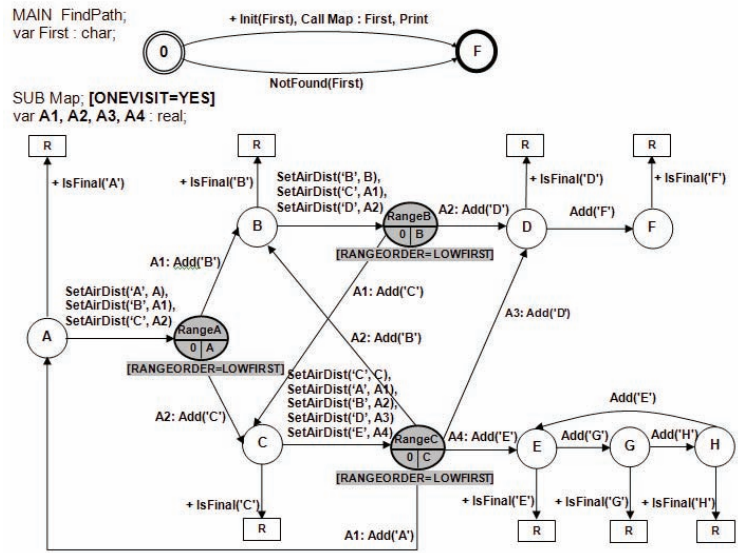

**Figure 5.** Non-recursive hill-climbing

In addition to being strategy-independent, the CN is also problem-independent - the same for different problems for which a given search strategy is applied (e.g., for the road map and 8-puzzle problems). As far as not only the CN but the complete CNP code is concerned, certain components must necessarily be problem-dependent. In our particular implementation (available at [5]), problem-dependent are a few help functions such as *arc_cost* and *h*. Of course, a few problem-dependent data structures will also be needed such as, for example, the representation of the map in the road map problem.

**2.2 Using strategy-dependent solutions**

One could also write separate (and probably more efficient) CNP implementations for a specific strategy. Even for such an algorithm the CN will remain unchanged. Simply, some of the primitives used must be adjusted to the particular strategy. Mainly, this will affect primitives *pushOpen* and *sortOpen*.

**2.3 Abstraction and expressiveness of CNP procedural implementation**

The CNP simulations of the search strategies discussed in Section 2 are simple to apprehend, design and implement but at the same time are rather general and abstract. (Well, one should expect similar characteristics from a CNP solution.)

It is worth mentioning that ideas similarly related to abstraction and generality can be found in well-designed object-oriented implementations of the search algorithms (e.g., [12]). These implementations use advanced object-oriented programming concepts such as abstract classes and interfaces. We would like to note that a CNP solution in a CNP environment with an object-oriented underlying language (environments such as *WinSpider* or *Spider#* - the latter being currently developed) could combine such typical advanced OOP possibilities with the possibilities offered by CNP itself.

The CNP solutions in Section 2, however, do not show the really great potential of CNP. The reason is the fact that the above implementations actually emulate procedural algorithms and, consequently, do not involve any non-determinism.

## 3. Non-Procedural Implementation of Local Search Strategies

In order to be able to create non-procedural search implementations we must make use of the built-in in CNP search mechanism which is an extended version of backtracking [1-3,6,7]. Then we will be able to use a descriptive-type CN, and it is the built-in interpreter's responsibility to "compute" the CN by finding a successful path from the initial to a final node.

Trivially, if the search strategy we want to model is backtracking (or we don't care what search strategy will be used) then there is no need to write a corresponding search procedure at all – we simply declaratively describe the problem and leave the inference to the interpreter. Eventually, the CNP programmer might want to use some of the static search control tools [3] for a better efficiency, to solve some problems related to non-termination, to specify other requirements such as the maximum number of the solutions required or the maximal length of the solution paths.

There are numerous other search strategies – both heuristic or improved uninformed ones - that have evolved from backtracking. Non-procedural implementations of some of them were described in [8] with complete codes shown at [5]: optimal search with cutting off insipid paths (branch-and-bound), hill climbing, irrevocable hill climbing, nearest neighbor search, version of beam search, stochastic hill climbing, first-choice hill climbing. In all these implementations we have made use of the so called tools for dynamic control of the computation – dynamic control system options and control states [7,8]. Applying such tools the CN programmer can easily model different modes of choice of which arrow emanating from the current node to attempt first (such as the seemingly best first, or randomly, or within a range of evaluations, or a restricted number of arrows, etc.). The programmer can also modify other parameters of the backtracking (e.g., forbid backtracking). Such modifications can be even performed dynamically (using variables whose values can be change during computation).

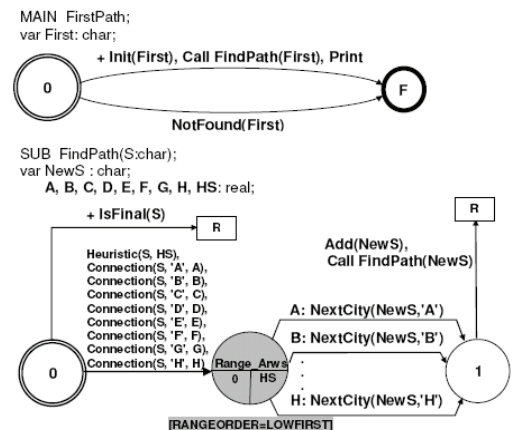

**Figure 6.** Recursive hill-climbing

Basically, a programmer can do anything they want with the arrows of the current state. The control may backtrack to a neighboring state which then becomes current. Therefore, we refer to this group of search strategies as "local search strategies". This phrase is related to but is not identical with the phrase "local search" as usually understood in the literature [10,13,14] where the emphasis is different. Note also, that our local search is usually revocable.

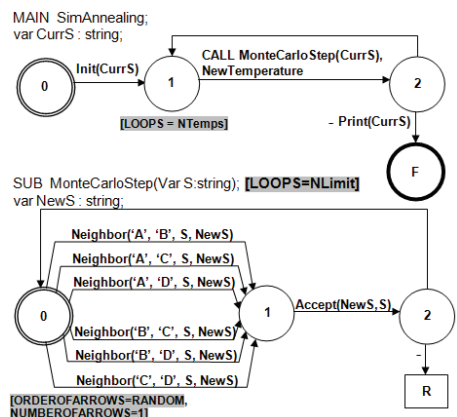

**Figure 7.** Simulated annealing

For more information on implementations of local search strategies the reader is referred to [7,8]. Only one example is shown here in Fig. 5 – one of the three discussed implementations of the hill climbing algorithm for a given road map that includes cities A through H.

When the state space is large, it is preferable to apply recursive CNP solutions. A recursive equivalent of the hill-climbing strategy for the road map problem is shown in Fig. 6. More details can be found in [8]. Note that the recursive solution remains non-procedural – hidden for the programmer backtracking is possible back through subnet calls.

## 4. Mixed Implementations

Not all local search algorithms, however, are suited for fully non-procedural CNP implementation. Some more complex strategies include elements that need procedural implementation. The resulting CN solution should be called a mixed implementation.

An example (from [8]) is shown in Fig. 7. It represents a solution to the Traveling Salesperson Problem (for cities A, B, C and D) using the Simulated annealing heuristic strategy. *MonteCarloStep* subnet defines declaratively a 'chunk' of the state space. The outer loop typical for the simulated annealing strategy, is modeled by the main subnet.

## 5. Conclusion

The report identifies, classifies and illustrates the ways in which search problems may be approached by a CNP programmer. Many local search strategies based on or derived from backtracking and hill-climbing can be implemented non-procedurally using the supported in CNP tools for computation control.

Non-local algorithms can be modeled using directly the presented generic procedural CNP solution or at least using the underlying pattern. Finally, more complex algorithms may require a mixed approach comprising elements of both types – procedural and non-procedural – at different levels.